\documentclass{elsart}

\usepackage{natbib}

 \usepackage{graphicx}

\usepackage{amssymb}

\newcommand\farcs{\mbox{$.\!\!^{\prime\prime}$}}%

\def\sax1808{SAX~J1808.4--3658}
\def\second{XTE~J1751--305}
\def\third{XTE~J0929--314}
\def\fourth{XTE~J1807--294}
\def\fifth{XTE~J1814--338}
\def\igr{IGR~J00291+5934}
\def\hete{HETE~J1900.1--2455}

\def\integral{{\it INTEGRAL}}
\def\amsp{{AMSPs}}

\def\nudot{\dot{\nu}}

\def\kte{kT_{\rm e}}

\def\rg{R_{\rm s}}
\def\Rin{R_{\rm in}}

\def\taut{\tau_{\rm T}}
\def\be{\begin{equation}}
\def\ee{\end{equation}}
\def\beq{\begin{eqnarray}}
\def\eeq{\end{eqnarray}}
\def\msun{{\rm M_\odot}}

\def\keV{{\rm keV}}

\def\Dop{\delta}

\begin{document}

\begin{frontmatter}



\title{Accretion-powered millisecond pulsars}


\author{Juri Poutanen}

\address{Astronomy Division, P.O.Box 3000, FIN-90014 University of Oulu, Finland}

\begin{abstract}
I review X-ray observations of accretion-powered millisecond pulsars and
current theories for formation of their spectra and pulse profiles.
\end{abstract}


\end{frontmatter}

\section{Introduction}
\label{sec:intro}

A number of rapidly spinning neutron stars in
low-mass X-ray binaries were discovered with the
{\it Rossi X-ray Timing Explorer} ({\it RXTE}) in the recent
years. These discoveries confirm the ideas on the formation
of radio (recycled) millisecond pulsars in low-mass X-ray binaries
(see review by \citealt{b95}).
Accretion of matter onto a neutron star results in its
spin increase to millisecond periods if the
magnetic field of the star is below about $10^9$ G.

Thirteen sources show nearly coherent oscillations
for a few seconds during X-ray bursts
at frequencies ranging from 270 to 619 Hz \citep[see][for a review]{sb03}.
These are now called  nuclear-powered millisecond pulsars.
The number of accretion-powered millisecond pulsar (AMSP)
showing pulsations in the persistent emission
reached seven by June 2005.
A more observationally inclined review of AMSP is given
by \citet{w05}. Here I concentrate on the results of the
X-ray spectroscopy, analysis of pulse profiles, and our present
theoretical understanding.

\section{Pulsars parameters}

\label{sec:param}

\begin{table}
\caption{
Parameters of the accretion-powered millisecond pulsars}
\vspace{12pt}
\begin{tabular}{llccccl}
\hline
& Source & $P_{\rm orb}$$^a$  & $\nu_{\rm spin}$$^b$ & $a_x \sin i$$^c$ & $f_x$$^d$   &  $M_{\rm c,min}$$^e$ \\
&        &  (min)           &  (Hz)           &  (lt-ms) & (M$_\odot$) & (M$_\odot$) \\
\hline
1 & SAX J1808.4$-$3658   & 121   & 401 &  62.809  & $3.779\times10^{-5}$  &  0.043  \\
2 & XTE J1751$-$305      &  42.4 & 435 &  10.113 & $1.278\times10^{-6}$  &  0.014 \\
3 & XTE J0929$-$314     &  43.6 & 185 &  6.290   & $2.9\times10^{-7}$    &  0.0083  \\
4 & XTE J1807$-$294      &  40.1 & 191 &   4.75   & $1.49\times10^{-7}$   &  0.0066 \\
5 & XTE J1814$-$338      & 257   & 314 &   390.3  & $2.016\times10^{-3}$  &  0.17  \\
6 & IGR J00291+5934    & 147   & 599 &  64.993  & $2.813\times10^{-5}$  &  0.039 \\
7 & HETE J1900.1$-$2455  &  83.3 & 377 &  18.39   & $2.00\times10^{-6}$   &  0.016 \\
\hline
\end{tabular}

\mbox{} \\
\noindent { $^a$ Orbital period; $^b$ neutron star spin frequency;
$^c$ projected semimajor axis;
$^d$ pulsar mass function; $^e$ minimum companion mass for a
$M_{\rm x}=1.4 {\rm M}_\odot$ neutron star.\\
References: [1] \citet{wk98a,cm98}; [2] \citet{ms02,mar02};
[3] \citet{rss02,gal02}; [4] \citet{mar03,k04,fal05a};
[5] \citet{ms03,mss03}; [6] \citet{eck04,mss04,mgc04,gal05};
[7] \citet{v05,mkv05,kmv05}.}
\label{tab:param}
\end{table}


The first real  AMSP \sax1808\  was discovered in 1998  by \citep{wk98a}.
Now (September 2005) there are seven AMSPs with spin frequencies
from 185 Hz up to 599 Hz (see Table 1).
The fastest AMSP, \igr\ with the period of just 1.67 ms
is the fifth fastest among all known pulsars
(including radio- and nuclear-powered MSPs).
The last AMSP  \hete\ was discovered in June 2005.

AMSPs show pulse frequency variations.
These observations  are  very important for
understanding of the evolution of the neutron stars in low-mass X-ray binaries
towards radio MSPs. They would also shed  some light
on a complicated problem of the interaction of the magnetosphere  with the
accretion flow.
One expects a spin-up rate
$$
\dot{\nu} = 3.7\times 10^{-13}\frac{L_{37} }{\eta_{-1} I_{45} }
\left( \frac{R_{\rm m}}{R_{\rm co} } \right) ^{1/2} \left( \frac{M}{1.4\msun}  \right) ^{2/3}
 \left( \frac{\nu_{\rm spin}}{600}  \right) ^{-1/3} \ \mbox{Hz s$^{-1}$} ,
$$
where (notation $Q=10^x Q_x$ in cgs units is used)
$I$ is the neutron star moment of inertia, $L$ is the
luminosity, $\eta$ is the accretion efficiency, $R_{\rm m}$ and $R_{\rm co}$ are the
magnetospheric and corotation radii.
Some reported $\dot{\nu}$ are, however, {\it negative} implying pulsar slowing down
during the outburst \citep{gal02,m03}.
\citet{mar04} finds
wild swings in the apparent spin frequency of both signs
resulting in the total fractional phase shift less than 0.15.
These could be, however, artifacts of the pulse
profile variations
(see Sect.~\ref{sec:lags}).
The reported positive $\dot{\nu}\sim 8\times 10^{-13}$ Hz/s
for \igr\ \citep{fal05b}
is larger than expected by a factor of 5 (since $L_{37}\sim 0.37$).
A positional error of $\sim0\farcs 7$ could result in
such a large derivative,  while the  radio position is known with a $0\farcs1$
accuracy \citep{rup04}. The motion of the radio source itself
does not produce an error larger than $\sim 0\farcs3$ (for the distance of 5 kpc)
a week after the outburst, confirming the reality of the pulsar spin-up and
possibly implying a small moment of inertia $I$.

Accretion-powered pulsars reside in very compact binary systems with
orbital periods ranging from 40 min to 4.3 h. Surprisingly three out of 7 pulsars
have $42\pm2$ min orbits. Adding to this set also 4U1626-67,
 4U1916-05 and X1832-330 (in globular cluster
NGC 6652) with orbital periods of 42, 50, and 44 min, respectively,
it becomes clear that this interesting fact deserves some explanation
\citep[see][for a possible scenario]{np03}.
The pulsar mass function
$$f_x=  (M_{\rm c}\sin i) ^3 / (M_{\rm c}+M_x) ^2   =
 4\pi^2 (a_x \sin i)^3 / G P_{\rm orb}^2$$
is very low for all the objects implying extremely low companion masses
consistent with degenerate white (helium or carbon-oxygen) or brown
dwarfs \citep{bc01,mar02,gal02,fal05a,gal05} except \fifth\ which contains
probably a hydrogen-rich star \citep{k05}.
All \amsp\ are transients with the outbursts repeating
every few years and lasting a few weeks. They have
a rather low
time-average accretion rate of $\sim 10^{-11} {\rm M}_\odot /{\rm yr}$,
which could be the main reason for magnetic field still to be strong
enough for pulsations to be observed \citep{czb01}.

\section{Broad-band X-ray spectra}

\begin{figure}
\centering \includegraphics[width=13 cm,height= 7 cm]{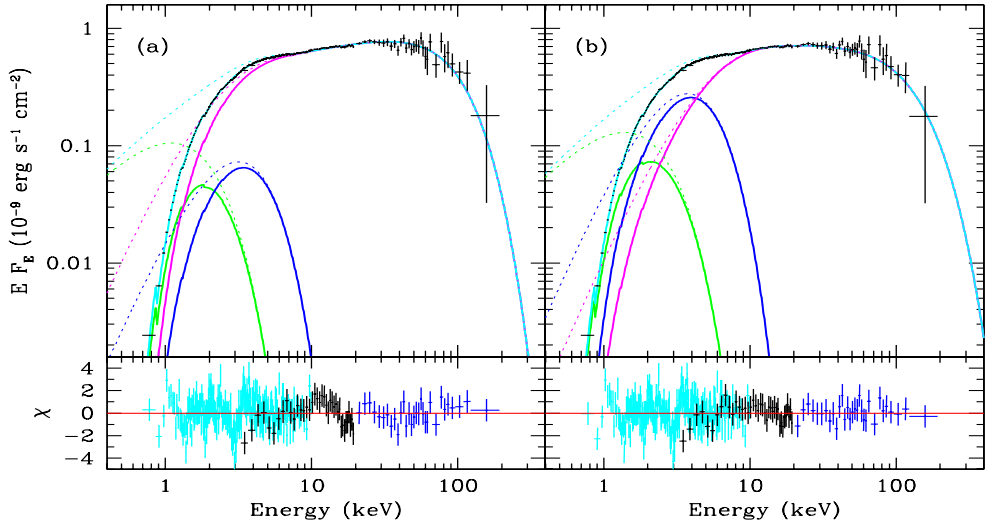}
  \caption{Spectrum of \second\ as observed by
{\it XMM}/EPIC-pn, {\it RXTE}/PCA  and HEXTE.
The models consists of the multicolour disc (peaking at $\sim 1$ keV),
single-temperature blackbody (peaking at $3-4$ keV), and thermal
Comptonization of the blackbody photons.
The dotted curves show unabsorbed spectral components.
Panel (a) shows a model where seed photons for Comptonization have
the same temperature as the blackbody photons, while
in panel (b) the seed photons are hotter.
The lower panels shows the residuals.
See  \citet{gp05} for details.}
  \label{fig:spectrum_1751}
\end{figure}

The broad-band coverage of the {\it RXTE} together with {\it XMM} gave a possibility
to study the spectra of AMSP in great details.
The spectra  can be modelled by three components: two soft, thermally looking
ones below a few keV and a power-law like tail (see Fig.~\ref{fig:spectrum_1751}).

The two soft components which can be
modelled as thermal emission from a
colder ($kT \sim 0.4-0.6$ keV) accretion disc and a hotter ($\sim 1$
keV) spot on the neutron star surface.
The softer components in \second\ and \fourth\
are studied with the {\it XMM} by \citet{gp05,fal05a}.
The obtained inner disk radius $\Rin\sim 10-15\ {\rm km}/\sqrt{\cos i}$
is consistent with the flow disrupted by the neutron star magnetosphere within
a couple of stellar radii.
The hotter black body normalization corresponding to
the area of $\sim30- 100\ {\rm km}^2$ and its pulsation
are consistent with it being
produces in a spot at the neutron star surface.

A power-law tail (having spectral photon index $\Gamma\sim1.8-2.1$)
shows a cutoff around 100 keV and
can be fitted with  thermal Comptonization.
The electron temperature of the Comptonizing medium
is around $\kte\sim$20--60 keV and Thomson optical depth of $\taut\sim$0.7--2.5
(for  a plane-parallel slab geometry) \citep{gdb02,pg03,gp05,fal05a}.
Weakness of the Compton reflection from the disk
  indicates that the solid angle covered by the disk as viewed
from the main emission source (accretion shock) is small,
being consistent with $\Rin\sim40$ km \citep{gp05}.

The broad-band X-ray spectra of \amsp\ are very similar to each other.
They also show very little variability during the outbursts
\citep[see e.g.][]{g98,gp05,fal05b}.
When fitting spectra with thermal Comptonization models,
one also finds that the product of the electron temperature
and optical depth is almost invariant (e.g.
 $[\kte, \taut]=[60\ \keV, 0.88]$ in \sax1808;
 $[33\ \keV, 1.7]$ in \second,
$[37\ \keV, 1.7]$ in \fourth,
$[49\ \keV, 1.12]$ in \igr; see \citealt*{gp05,fal05a,fal05b}).
The constancy of the spectral slope can be used as an argument that the
emission region geometry does not vary much with the accretion rate.
If the  energy dissipation takes place in a hot shock, while the cooling
of the electrons is determined by the reprocessing of the hard
X-ray radiation at the neutron star surface (two-phase model, e.g.
\citealt{hm93,ps96,mbp01}),
the spectral slope is determined by the energy balance in the hot phase
and, therefore, by the geometry.

\section{Phase-resolved spectroscopy, pulse profiles, and time lags}


\label{sec:lags}

\begin{figure}
\centering \includegraphics[width=10 true cm]{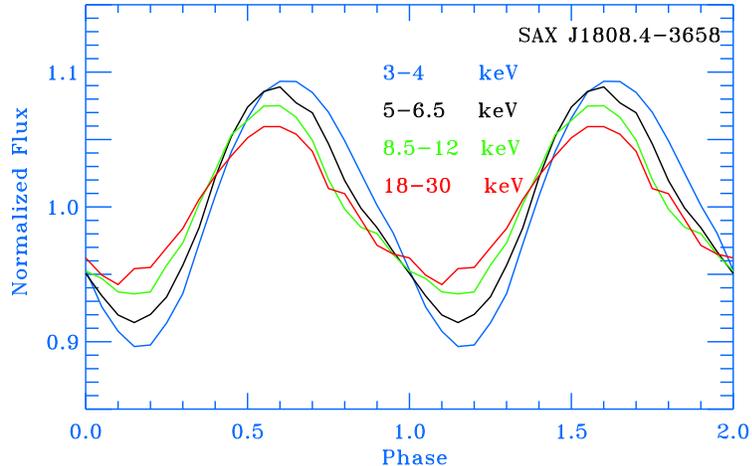}
\caption{Normalized pulse profiles of \sax1808\ at different energies
during the April 1998 outburst. } 
\label{fig:data_sax1808}
\end{figure}

The pulse profiles from \amsp\ are rather close to sinusoidal with
peak-to-peak oscillation amplitude
$A=(F_{\max}-F_{\min})/(F_{\max}+F_{\min})$
between 4   (in \second)  and 12 per cent (in \fifth).
At energies above 50 keV the \integral\ data on \igr\ show the
increase of the pulse amplitude up to $\sim$ 25 per cent \citep{fal05b}.
Deviations from the sine wave are stronger at higher energies
(see Fig.~\ref{fig:data_sax1808}). The harmonic content also
is stronger when $A$ is larger (e.g.
the   harmonic-to-fundamental ratio is $\sim 0.03$ in \second\ \citep{gp05},
while it is $\sim 0.33$ in \fifth\ \citep{sms03}.

\begin{figure}
\centering
 \includegraphics[width=6 true cm,height= 5.2 true cm]{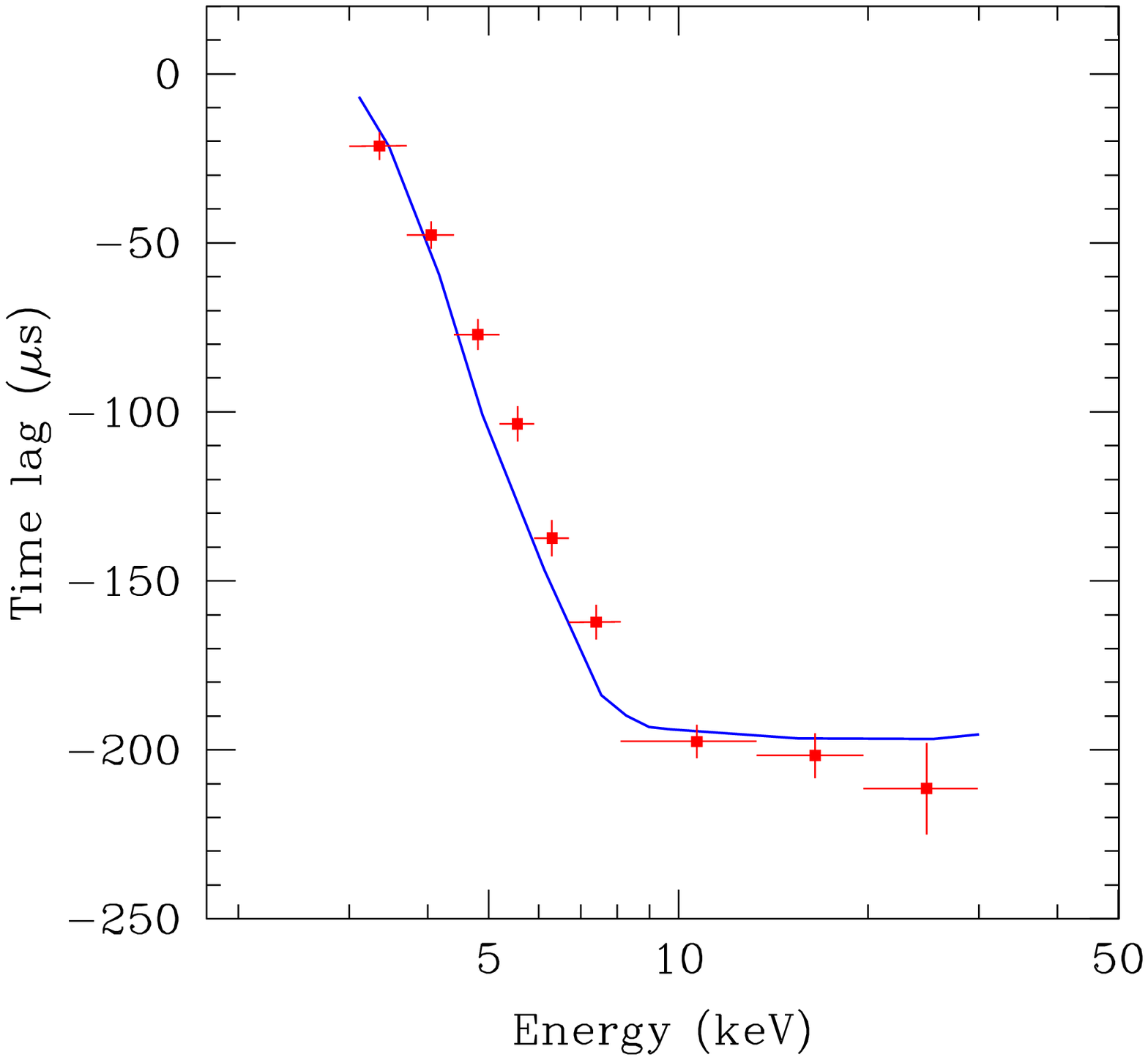}
 \includegraphics[width=7 true cm,height= 5.2 true cm]{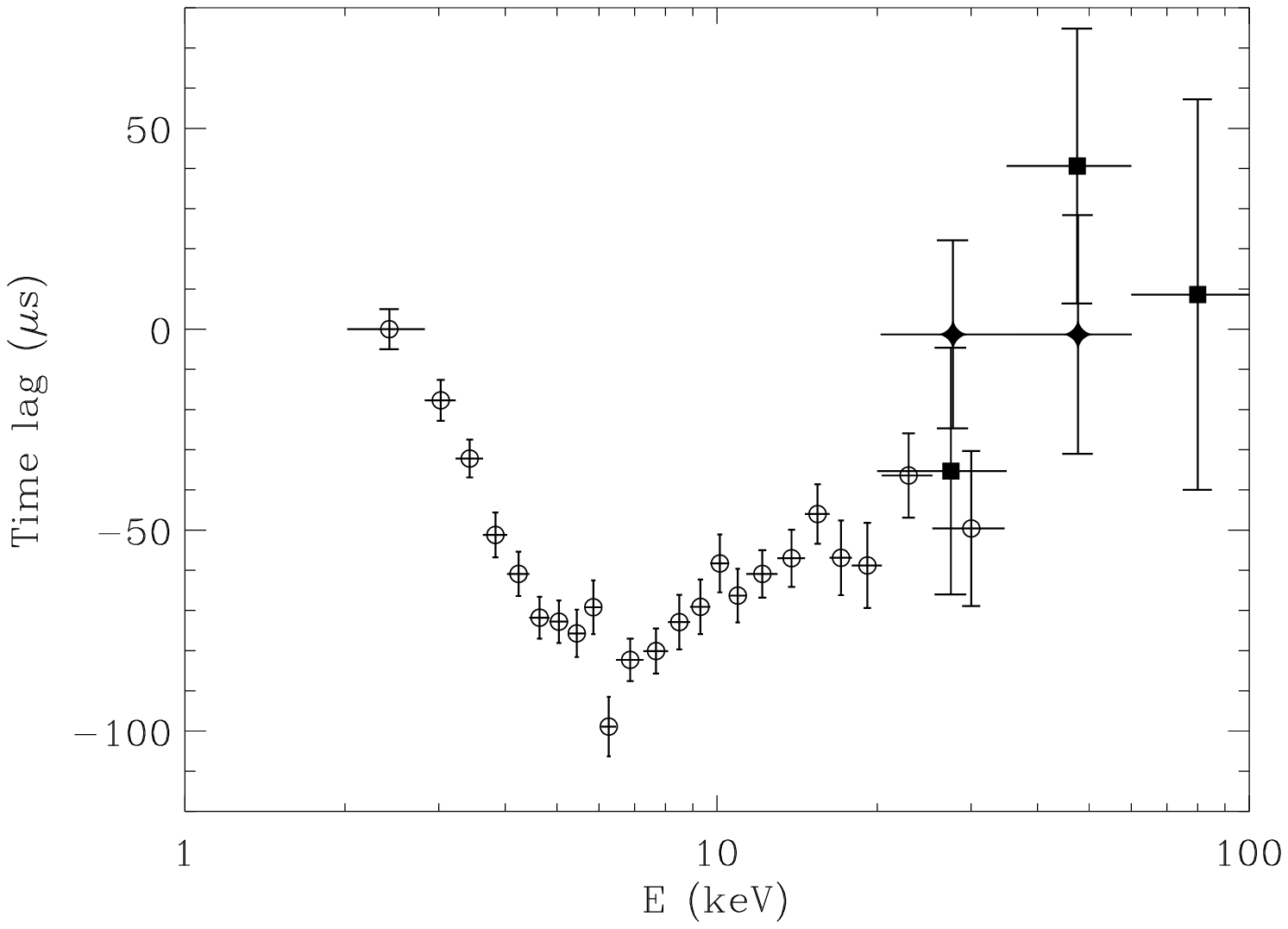}
\caption{Time lags as a function of energy
for \sax1808\ (left panel;   \citealt{gdb02}) and
\igr\ (right panel;   \citealt{fal05b}). }
\label{fig:time_lags}
\end{figure}

Pulse profiles at higher energies reach their peaks at an earlier phase
relative to the soft photons resulting in the soft time lags.
In \sax1808\ \citep{cmt98,gdb02}, \second\ \citep{gp05}, and \third\
\citep{gal02}
the lags increase (in absolute value) with energy up to about 7--10 keV after they
saturate (see Fig.~\ref{fig:time_lags}, left panel).
In \igr\ (see Fig.~\ref{fig:time_lags}, right panel; \citealt{gal05,fal05b}),
the behaviour is more complicated and
the lags decrease between 7 and 15 keV, seemingly saturating
at higher energies.

\begin{figure}
\centering \includegraphics[width=11 true cm,height= 5.2 true cm]{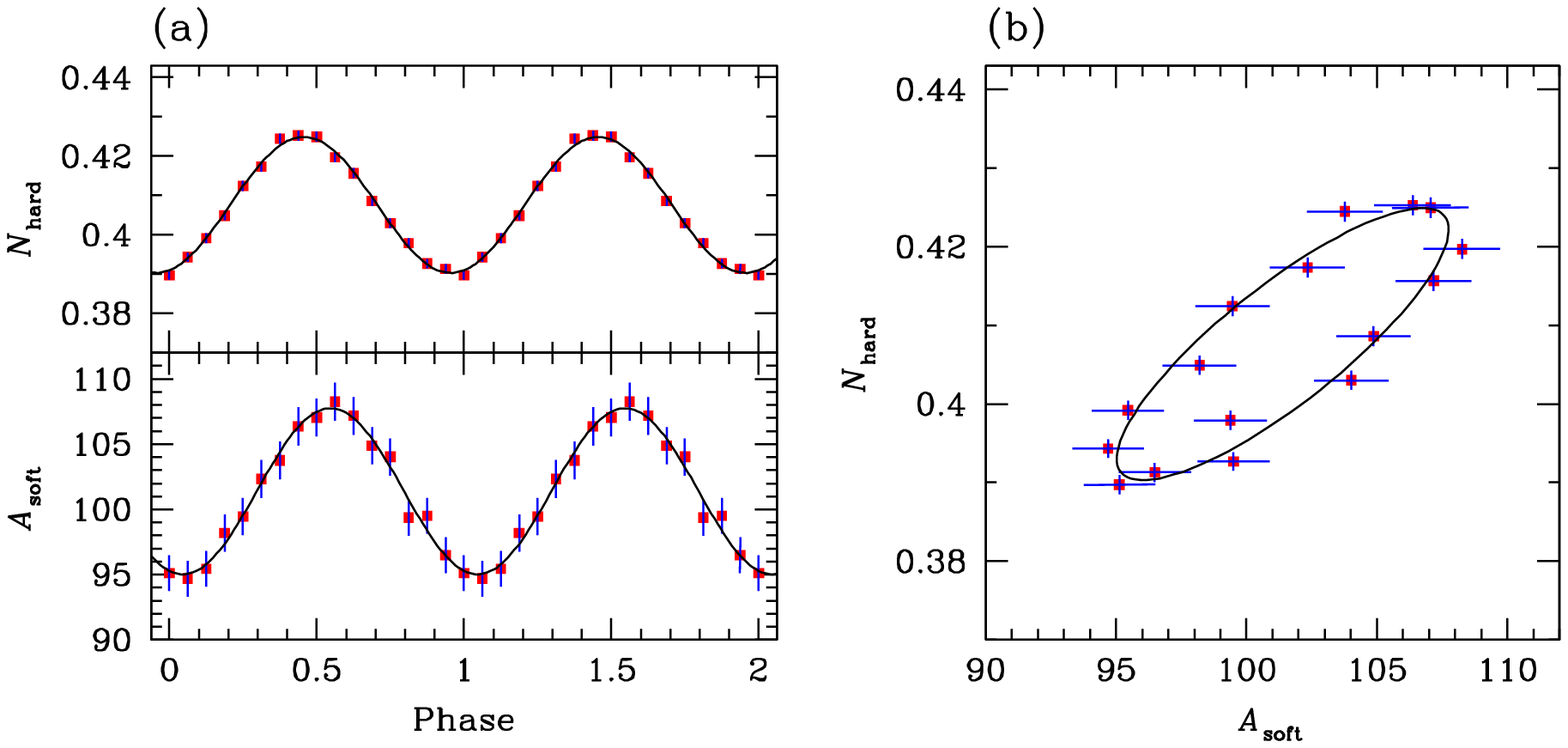}
\caption{Pulsation  of the two model components, blackbody and
Comptonization, in \second. From \citet{gp05}.}
\label{fig:fars_bbtf}
\end{figure}

One can point out that the  contribution of the black body decreases
exponentially with energy and the lags increase significantly at the same time.
Above $\sim 7$ keV, the blackbody's contribution is negligible and the
lags saturate. When fitting phase-resolved spectra with a two-component
model (blackbody + Comptonization), one clearly sees that the normalizations of the components
do not vary in phase, with black body lagging the Comptonized emission
(see Fig.~\ref{fig:fars_bbtf} and \citealt*{gdb02,gp05}). In \sax1808\ the
profile corresponding to the Comptonized emission is clearly non-sinusoidal.
Difference in profiles of the two components can be explained by
their different emission pattern \cite{pg03}. Its variation
may cause pulse shape  change  resulting in the swings of
$\nudot$ discussed in Sect.~\ref{sec:param}.

\section{Modelling the pulse profiles of \amsp}

One believes that the bulk of the X-ray emission
observed from \amsp\  originates from polar caps where
the gas stream channels by the neutron star magnetic field impacts the
stellar surface forming a shock.
This is supported by a weak energy dependence of the variability amplitude
and a  fairly constant spectral shape  as a function of
pulse phase (any additional source of radiation would have to have
a spectrum identical to that of the shock),
as well as the observed broadening of the pulse peak
in the power-spectrum of \sax1808 due to modulation of the aperiodic variability
by the spin period \citep{men03}.

\begin{figure}
\centering \includegraphics[width=9.5 true cm]{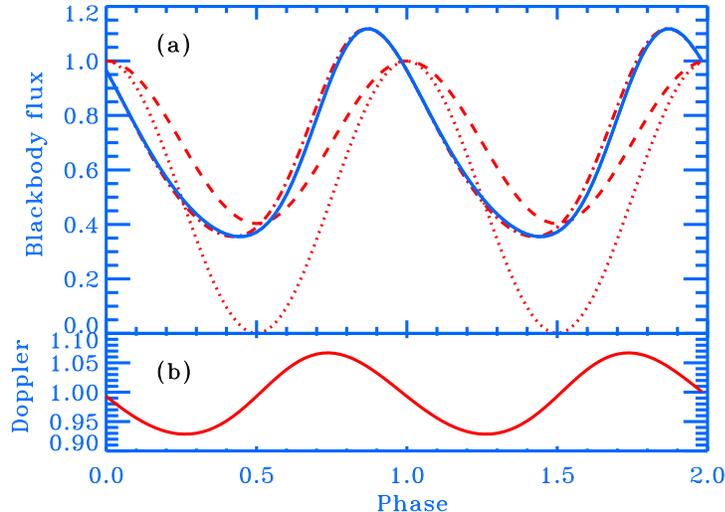}
  \caption{Pulse profile for a small black body spot
  at the neutron star surface.
  (a) Dotted curve shows the  profile for a
  slowly rotating star in Newtonian approximation.
  Gravitational light bending reduces variability amplitude (dashed curve).
  Doppler effect due to rapid rotation  skews the profile
  (dot-dashed curves). Accounting for the light travel delays
  further modifies the profile slightly (solid curve). (b) Doppler factor
  as a function of phase.
  We assumed neutron star mass $M=1.4\msun$ and radius  $R=10.3$ km,
  rotational frequency $\nu=600$ Hz, the
  inclination $i=45^{\rm o}$, and
  the polar angle of the spot center $\theta=45^{\rm o}$.
   }
   \label{fig:relat}
\end{figure}

\subsection{Computing pulse profiles}

Pulse profile shape and variability amplitude
carry the information about the
compactness of the neutron star, spot size and its position at the star, and the
emission pattern.
Gravitational light bending tends to decrease the pulse amplitude
\citep{pfc83,rm88,page95,ll95,b02}.
An earlier attempt to incorporate the effect of rapid rotation using Kerr metric
\citep{cs89} did not include the dominant Doppler boost caused by the star's surface
motion in the locally non-rotating frame. This
was criticized by \citet{brr00}. They also noticed that
Doppler boost and light travel time delays are important for
shaping the profiles, while the frame dragging
affects them at a $\sim1\%$ level. The Monte-Carlo simulations used by
\citet{brr00} for the light curve calculation are, however, extremely
time-consuming and are hardly suitable for data fitting.

As frame dragging is not so important, one can consider
Schwarzschild metric instead. One can account for Doppler effect
making Lorentz transformation from the frame rotating with the star to a
non-rotating frame and then follow light trajectory in  the Schwarzschild space-time
to infinity (``Schwarzschild + Doppler" or SD approach).
An attempt to formulate the problem was done by \citet{ml98} who
considered  the spot and the observer  in the equatorial plane only.
Oscillation waveforms and amplitudes for arbitrary spot
and observer position were computed by \citet{wml01} and \citet{moc02}.
These paper provided no details on how to account for effect of aberration
on intensity and projected spot area.
A  formalism for computing the
pulse profiles accounting for Doppler boosting, relativistic aberration
and gravitational bending appeared for the first time in \citet{pg03}.
Effects of the anisotropy of Comptonized radiation from a shock
on the light curves and polarization   were studied by   \citet{vp04}.
Analytical formulae for oscillation amplitudes
are presented by \citet{p04}.
Light curves from realistic spots produced by accretion onto inclined magnetic
dipole are computed by \citet{kr05} using formalism of \citet{pg03,vp04,b02}.
Accuracy of the SD approach is discussed by \citet{clm05}.

\subsection{Main effects}

Let us now briefly describe the main effects
shaping the pulse profiles of \amsp. Without general or special
relativistic effects,
a small black body spot would produce sinusoidal variations (with possible
eclipses)
due to a change of the projected area. Light bending reduces the
variability amplitude (compare dotted and dashed
curves in Fig.~\ref{fig:relat}), while the pulse remains
almost sinusoidal \citep{b02}.
Relativistic aberration and Doppler boosting modify the observed
flux for rapidly  spinning  star. The projected area
is changed by the Doppler factor $\Dop$ due to aberration, and
the frequency-integrated specific intensity is multiplied by $\Dop^4$ 
\citep[see][]{pg03}. Since $\Dop$ reaches the maximum  a quarter of the period earlier than
the projected area (Fig.~\ref{fig:relat}b), the pulse becomes skewed to the left
(see Fig.~\ref{fig:relat} and compare it to Fig.~\ref{fig:data_sax1808}).
Light travel time delays  slightly modify the profile further (note,
that time delays are already accounted for in the flux calculations by
one of the Doppler factors).

\subsection{Oscillation amplitudes}

Due to the combined action of the Doppler effect and time delays,
the pulse profile deviates from a simple sine wave. (We consider
below the situation where the spot is always visible.)
The relative amplitudes (in the bolometric signal)
of the harmonic $A_1$ to that of the fundamental $A_0$
grows approximately  linearly with the rotational frequency \citep{p04}:
$$
\frac{A_1}{A_0}\approx \frac{5}{2}\ \frac{2\pi R}{c}\ \nu_{\rm spin} \ \sin i\ \sin\theta =
 0.16 \ \frac{R}{10\ {\rm km}} \ \frac{\nu_{\rm spin}}{300\ {\rm Hz}}  \ \sin i\ \sin\theta ,
$$
where $i$ is the inclination   and $\theta$ is
  the polar angle of the spot center.
Even a  slight deviation from the isotropic emission (e.g. in the form
$I(\alpha)\propto 1+b\cos\alpha$, where $\alpha$ is the   angle
from the spot normal and  $b$ is the anisotropy parameter)
leads to a  high  harmonic content \citep{pg03,p04,vp04}:
$$
\frac{A_1}{A_0}=\frac{b (1-\rg/R)/2\ \
\sin i\ \sin\theta }{1+2b [\rg/R+(1-\rg/R) \cos i\ \cos\theta]},
$$
where $\rg=2GM/c^2$.
The harmonic content $A_1/A_0$ as well as the total rms
are proportional to $\sin i\ \sin\theta$ \citep{b02,pg03,pg04,p04}.
Therefore, one expects harmonics to be relatively stronger when
rms is larger which is exactly what is observed.
Increasing the spot size leads  to a reduction of the
amplitude of the harmonic first and of the total rms later \citep{wml01,p04}.

If the spectrum has a sharp cutoff, the rms amplitude of
the pulse at energies above the cutoff
increases dramatically. The flux   varies as  $\propto\Dop^{3+\Gamma}$
 \citep{pg03,vp04}, where
$\Gamma (E) = 1 - {\rm d} \ln F_E / {\rm d} \ln E$ is
 the energy-dependent photon spectral index,
 resulting in the rms energy dependence $\propto 3+\Gamma(E)$.
For the spectrum in the form
$F_E \propto E^{-(\Gamma_0-1)} \exp( -[E/E_{\rm c}]^{\beta})$, which
is typical for Comptonization, the photon index $\Gamma(E)=
\Gamma_0 + \beta (E/E_{\rm c})^{\beta}$.
At low energies, $\Gamma\approx\Gamma_0$, and
rms  is a very weak function of energy.
Close to the cutoff, the spectral index rapidly increases
resulting in a significant grow of the rms, as
observed in the \integral\ data for \igr\ \citep{fal05b}.
In case of coherent oscillations observed during
X-ray bursts, a linear increase of the rms with energy is expected, since
the spectrum is close to a blackbody which has an exponential cutoff
(see also \citealt*{ml98,moc03}).

\begin{figure}
\centering 
\includegraphics[width=10cm]{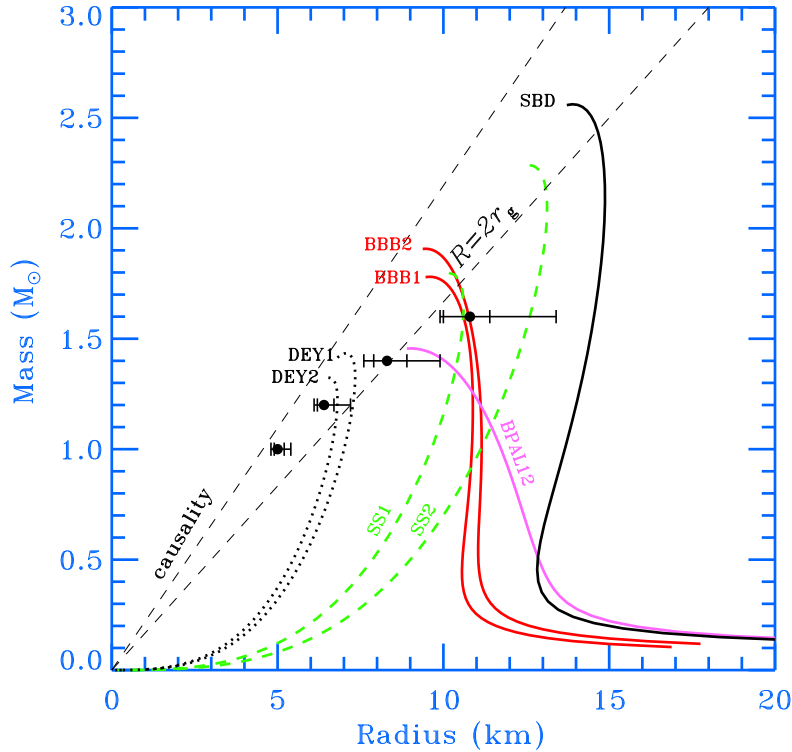}
  \caption{Constraints on the neutron star mass-radius relation obtained
  by fitting the pulse profiles of
  \sax1808\  (filled circles with error bars) together with a set
  of equations of state for neutron and strange stars.
  From \citet{pg03}.}
\label{fig:eos}
\end{figure}

\subsection{Constraints on the neutron star equation of state and spot parameters}

The amplitude of  variability  and the
pulse shape (harmonic content)
can be used to put  constraints on the  compactness  of the neutron star,
position, size and the emission  pattern of a radiating spot.
In case of the coherent oscillations
observed during  the X-ray bursts the constraints on the
compactness are not very tight \citep{ml98,nss02,bsmm05}.
A very weak harmonic content
led to the conclusion that the
bright region (or the line of sight) must be near the rotational pole or
must cover half of the star \citep{moc02}.
Fitting the pulse profile shape (but not
the amplitude) of the oscillations observed during
the X-ray burst of \fifth\ was performed by \citet{bsmm05}, who
obtained  the preferred spot polar angle
$\theta=90^{\rm o}\pm 30^{\rm o}$.
This is highly improbable since an identical (in shape and phase)
pulse observed in the persistent emission
\citep{sms03} would require the magnetic poles to lie at the
equator and the antipodal spot should then be  visible too.

For \amsp\ the statistics in the pulse profiles is much better since
the pulse is folded over a longer observational period
(days or weeks rather than seconds as in the case of X-ray bursts).
This allowed to obtain tight constraints on the neutron star
compactness in \sax1808\ as well as the
emission pattern (consistent with Comptonization from the plane-parallel
shock), inclination,  and the spot polar angle  
(see \citealt*{pg03}, and Fig.~\ref{fig:eos}).

\section{Summary}

We have reviewed here X-ray observations of the AMSPs concentrating
on the broad-band spectra and the energy-dependent pulse profiles.
Recent advances in the theoretical modelling of the spectral
and timing characteristics are also described.
At the observational side, we are missing the details of the pulse
profile and spectral evolution during the outbursts. One would
expect that the changing accretion rate would force a change in the accretion
pattern on to a neutron star surface, causing corresponding profile variations.
Detailed studies of the reflection amplitude and
frequencies of quasi-periodic oscillations should
reveal changes in the inner disk radius during outbursts.
The pulse profile should also react correspondingly as the visibility of
the antipodal spot depends on the disk radius.
Combined with the spin-up rate observations, these would shed some light
on a complicated problem of the magnetosphere interaction with the accretion flow.
On the theoretical side, we need a detailed model
describing the dynamics of the accretion flow onto the neutron
star surface including radiation feedback. It is not clear yet what
fraction of the material follows the magnetic field lines towards the poles.
The impact of the realistic spot geometry
and the radiation pattern on the pulse profiles and the
derived parameters should be studied further.

\section*{Acknowledgements}

This work is supported by the Academy of Finland, the
Jenny and Antti Wihuri Foundation, the
Vilho, Yrj\"o and Kalle V\"ais\"al\"a Foundation.




\end{document}